\documentclass[12pt,A4paper]{article}
\usepackage{amsthm,amsmath,natbib,amssymb,amsfonts,bm}
\usepackage{graphicx}
\usepackage{epstopdf}
\usepackage{placeins}
\def\hang{\hangindent\parindent}
\def\rf{\par\noindent\hang}

\def\rf{\par\noindent\hang}

\newtheorem{theorem}{Theorem}

\newtheorem{lemma}{Lemma}

\newcommand{\bvarepsilon}{\boldsymbol{\varepsilon}}
\newcommand{\by}{\boldsymbol{y}}
\newcommand{\bX}{\boldsymbol{X}}

\newcommand{\bzero}{\boldsymbol{0}}

\newcommand{\bbeta}{\boldsymbol{\beta}}
\newcommand{\blambda}{\boldsymbol{\lambda}}

\newcommand{\sd}{\text{sd}}
\newcommand{\var}{\text{var}}
\newcommand{\corr}{\text{corr}}

\begin{document}

\baselineskip=21pt

\begin{center} {\Large{\textbf {Confidence intervals centered on bootstrap smoothed estimators}}}
\end{center}


\begin{center}
\large{\textbf{Paul Kabaila}$^*$  and \textbf{Christeen Wijethunga}}
\end{center}

\phantom{1}
\vspace{-1.3cm}

\begin{center}
Department of Mathematics and Statistics, La Trobe University, Melbourne   Victoria, Australia.
\end{center}

\noindent ABSTRACT

\smallskip

\noindent 
Bootstrap smoothed (bagged) parameter estimators have been proposed as an improvement on estimators found after preliminary data-based model selection. 
The key result
of Efron (2014) is a very convenient and widely applicable formula for 
a delta method approximation to the standard deviation of the bootstrap smoothed estimator. 
This approximation 
provides an easily computed guide to the accuracy of this estimator. 
In addition,  Efron (2014) proposed a confidence interval centered on the bootstrap
smoothed estimator, with width proportional to the estimate of this approximation to the standard deviation. 
We evaluate this confidence interval in the 
scenario
of two nested 
linear regression models, the full model and a simpler model, and a preliminary test
of the 
null hypothesis that the simpler model is correct. 
We derive computationally convenient
expressions for the ideal bootstrap smoothed estimator and the coverage probability and 
expected length of this confidence interval. 
In terms of coverage probability, this confidence interval outperforms
the post-model-selection confidence interval with the same nominal coverage and based on the same preliminary test. 
We also compare the performance of confidence interval centered on the bootstrap smoothed estimator, in terms of expected length, to the usual confidence interval, with the same minimum coverage probablility, based on the full model.

\bigskip


\bigskip

\noindent { * Corresponding author address: Department of Mathematics and Statistics, La Trobe University, Victoria 3086, Australia.
	Tel: +61-03-9479-2594; fax: +61-03-9479-2466. E-mail address: P.Kabaila@latrobe.edu.au}

\newpage


\noindent {\bf 1. Introduction}

\medskip

In applied statistics it is common practice to carry out preliminary data-based model selection
(using e.g. hypothesis tests or minimizing a criterion such as AIC) and then to use the selected model to carry out further inference for the parameter of interest on the assumption that the selected model had been given to us
{\sl a priori}, as the true model. We refer to such further inferences as 
post-model-selection inferences. 
Post-model-selection point estimators have the inherently undesirable property that they 
are discontinuous functions of the data. In the terminology of Efron (2014),
they are ``jumpy''.
Bootstrap smoothed (or bagged, Breiman, 1996)  estimators have been proposed as an
improvement on post-model-selection estimators. Bootstrap smoothed estimators are 
smoothed versions of the post-model-selection estimator. 
The key result of Efron (2014) is a new formula for a delta method approximation to the standard deviation of the bootstrap smoothed estimator. 
This formula is valid for any exponential family of models and has the attractive feature that it simply re-uses
the parametric bootstrap replications that were employed to find this estimator.
It also has the attractive feature that it is applicable in the context of complicated data-based
model selection. This formula provides an easily computed guide to 
the accuracy of the bootstrap smoothed estimator.

Post-model-selection confidence intervals have the inherently undesirable 
property that they have endpoints that are discontinuous functions of the data.
Furthermore, these confidence intervals may have minimum coverage probability far below nominal (see e.g. Leeb and P\"otscher, 2005 and Kabaila, 2009). Confidence intervals
that deal properly with  the ``model uncertainty" commonly encountered in applications
are desperately needed by statistical practitioners. Such confidence intervals should 
have (a) endpoints that are smooth functions of the data, (b) have the desired minimum coverage probability and (c) attractive expected length properties.

In response to this need, a number of frequentist model averaged confidence intervals 
have been proposed (Buckland \textit{et al.}, 1997, Hjort and Claeskens, 2003, Fletcher and Turek, 2011, Turek  and Fletcher, 2012). A related approach is 
the proposal of 
Efron (2014) of a confidence interval (CI) centered on the bootstrap
smoothed estimator.
This CI, with 
nominal coverage $1 - \alpha$, has half-width equal to the $1 - \alpha/2$ quantile of the standard normal distribution multiplied by the estimate of the delta
method approximation, ${\tt sd}_{\rm delta}\,$, to the
standard deviation of this estimator. 
We call this interval the 
${\tt sd}_{\rm delta}\,{\tt interval}$.

Wang {\it et al} (2014) assess the ${\tt sd}_{\rm delta}\,{\tt interval}$ using simulations to estimate weighted averages over values of the explanatory variables of the coverage, center and length of this CI. In terms of these weighted averages, this CI seems to perform well for the scenarios that they consider. However, these weighted averages over the explanatory variables will tend to mask particular values of the explanatory variables for which the coverage is low or the expected length is large.

To rigorously evaluate the 
${\tt sd}_{\rm delta}\,{\tt interval}$, we consider the simple, though informative, scenario
of two nested normal linear regression models 
and parameter of interest 
$\theta$ a specified linear combination of the regression parameters.
These two nested models are the full model and the simpler model where $\tau$,
a distinct specified linear combination of the regression parameters, is set to 0.
This scenario was used by Kabaila, Welsh and Abeysekera (2016) 
and Kabaila, Welsh and Mainzer (2017)
to evaluate the frequentist model averaged confidence intervals proposed by 
Fletcher and Turek (2011) and Turek  and Fletcher (2012).
The bootstrap smoothed estimator that we consider is a smoothed version of the post-model-selection estimator obtained after a preliminary test of the null 
hypothesis that $\tau = 0$ against the alternative hypothesis that 
$\tau \ne 0$.



In Section 3, for this simple scenario of two nested regression models,
we derive a computationally convenient exact expressions for the ideal (i.e. in the limit as the number of bootstrap
simulations approaches infinity) bootstrap estimator.
The delta-method approximation ${\tt sd}_{\rm delta}$
to this standard deviation can be found using the formula of Efron (2014).

Let $\widehat{\theta}$ denote the least squares estimator of
$\theta$ (based on the full model). The usual CI based on the full model
is, of course, centered on $\widehat{\theta}$.
Also let $\widehat{\tau}$ denote the least squares estimator of
$\tau$ (based on the full model).
In Section 4, we consider the coverage probability of 
the ${\tt sd}_{\rm delta}\,{\tt interval}$.
 We show that this coverage probability
is determined by
 the known correlation
 $\rho = \corr(\widehat{\theta}, \widehat{\tau})$ and the unknown parameter
 $\gamma = \tau \big / \text{(standard deviation of $\widehat{\tau}$)}$. 
 We also show that this coverage probability is an even function of $\gamma$, for
every given $\rho$, and an even function of $\rho$, for every given $\gamma$.
We are therefore able to encapsulate the coverage probability function of 
the ${\tt sd}_{\rm delta}\,{\tt interval}$, for all possible choices of design matrix,  
parameter of interest $\theta$ and parameter $\tau$ that specifies the simpler model, using only the two 
parameters $|\rho|$
and $|\gamma|$.
An immediate consequence of the results of 
Section 3 is that when $\rho = 0$, 
the 
${\tt sd}_{\rm delta}\,{\tt interval}$ are identical to the 
usual CI, with actual coverage $1 - \alpha$, based on the full model. However, 
as $|\rho|$ increases the latter confidence interval increasingly differs from the
${\tt sd}_{\rm delta}\,{\tt interval}$.

Figure 1 shows the graph (solid line) of the coverage probability of the 
${\tt sd}_{\rm delta}\,{\tt interval}$
centered on the bootstrap smoothed estimator based on the post-model-selection estimator obtained after a preliminary hypothesis test,  with size 0.1, of the null hypothesis that the simpler model is correct.  This CI has
nominal coverage 0.95.
We consider $|\rho| =  0.7$.
Also shown in this figure is the graph (dashed line) of the coverage 
probability of the post-model-selection CI with the same nominal coverage 
and based on the same preliminary test. This panel provides an illustration of the fact, established through an extensive numerical investigation described in the Supplementary material, that 
the 
${\tt sd}_{\rm delta}\,{\tt interval}$ outperforms the post-model-selection CI, with the same nominal coverage 
and based on the same preliminary test, in terms of minimum coverage probability.

\FloatBarrier

\begin{figure}[h]
	\centering
	\begin{tabular}{c}
		\includegraphics[width=0.77\textwidth]{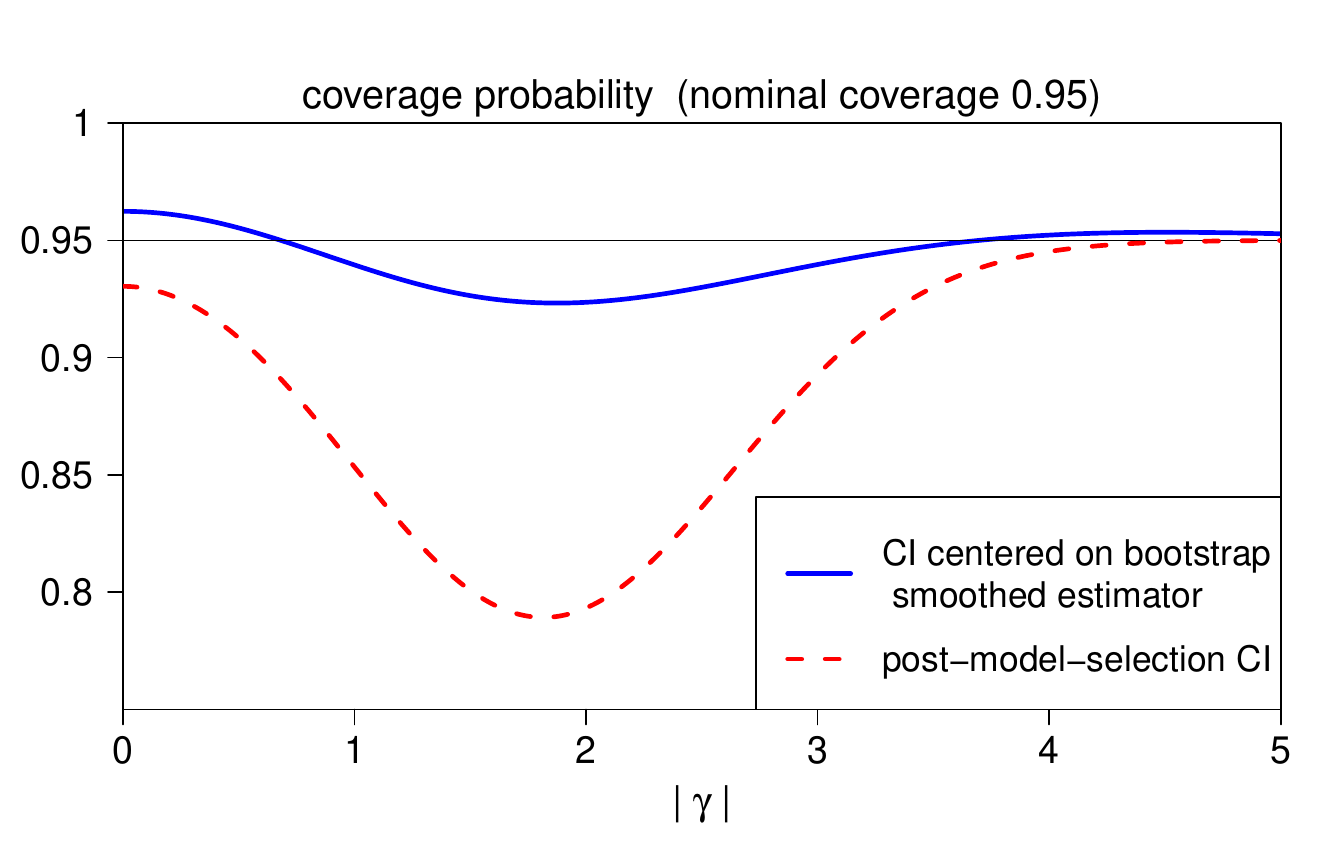}  
	\end{tabular}
	\caption{This figure shows a graph (dashed line) of the coverage probability of the post-model-selection CI, based on 
		a preliminary hypothesis test,  with size 0.1, of the null hypothesis that the simpler model is correct. This CI has
		nominal coverage 0.95. It also shows a graph (solid line) of the coverage probability for the
		${\tt sd}_{\rm delta}\,{\tt interval}$,
		based on the post-model-selection estimator obtained after the same preliminary test. This CI also has nominal coverage 0.95.
		Here $|\rho| =  0.7$.}
	\label{CP_plot_std}
\end{figure}

\FloatBarrier

A further measure of the quality of the
${\tt sd}_{\rm delta}\,{\tt interval}$ is its 
scaled expected length, 
where the scaling is with respect to the expected length
of the usual CI, with the
same minimum coverage probability, based on the full model.
In Section 5, we derive a computationally convenient formula 
for  the scaled expected length of 
the 
${\tt sd}_{\rm delta}\,{\tt interval}$. Using this formula, we provide a detailed examination of the scaled expected length properties of the 
${\tt sd}_{\rm delta}\,{\tt interval}$.

\bigskip

\noindent {\bf 2. The two models and the post-model-selection estimator}

\medskip

We consider two nested linear regression models: the full model
${\cal M}_2$ and the submodel ${\cal M}_1$.
Suppose that the full model ${\cal M}_2$ is given by
\begin{equation}
\label{SimplerScenario}
\by = \bX \bbeta + \bvarepsilon
\end{equation}
%
where $\by$ is a random $n$-vector of responses, $\bX$ is a known $n \times p$ matrix with linearly independent columns 
($p < n$), $\bbeta$ is an unknown $p$-vector of parameters and $\bvarepsilon \sim N(\bzero, \sigma^2 \boldsymbol{I}_n)$ with $\sigma^2$ known.
Suppose that 
$\bbeta = [\theta, \tau, \boldsymbol\lambda^\top ]^\top$, where $\theta$ is the scalar parameter of interest, 
$\tau$ is a scalar parameter used in specifying the model ${\cal M}_1$ and $\blambda$ is a ($p-2$)-dimensional parameter vector. The model ${\cal M}_1$ is ${\cal M}_2$ with $\tau = 0$. As shown in the Supplementary material, this scenario can be obtained by a change of parametrization from a more
general scenario.

We assume that the error variance $\sigma^2$ is known, as does Efron (2014, Section 4)
when he uses a linear regression model for the supernova data.
It is highly plausible that for a linear regression model, the known $\sigma^2$ case provides a good approximation to the case that $\sigma^2$ is
unknown, so that it must be estimated, and   
$n - p$ is reasonably large.

Let $\widehat{\bbeta}$ denote the least squares estimator of $\bbeta$, so that
$\widehat{\bbeta} = (\bX^\top \bX)^{-1} \bX^\top \by$.
Also let $\widehat{\theta}$ and $\widehat{\tau}$ denote the first and second components of
$\widehat{\bbeta}$, respectively.
Let $v_{\theta} = \var(\widehat{\theta}) / \sigma^2$, 
$v_\tau = \var(\widehat{\tau}) / \sigma^2$ and
$\rho=\corr(\widehat{\theta}, \widehat\tau)$. Note that $v_{\theta}$, $v_\tau$
and $\rho$ are known.
Let $\gamma = \tau / \big(\sigma v_\tau^{1/2} \big)$, which is an unknown parameter, and also
let  $\widehat{\gamma} = \widehat{\tau} / \big(\sigma v_\tau^{1/2}\big)$. 
We will express all
quantities of interest in terms of the random vector 
$\big(\widehat{\theta}, \widehat{\gamma} \big)$, which has a bivariate normal 
distribution with mean $(\theta, \gamma)$ and
known covariance matrix.

Suppose that we carry out a preliminary test, of size $\widetilde{\alpha}$, of the null hypothesis $\tau = 0$ against the alternative hypothesis
$\tau \neq 0$. The test statistic is $|\widehat{\gamma}|$, which has
the same distribution as $|Z|$, for $Z \sim N(0,1)$, under the null hypothesis.
Let the quantile $z_a$ be defined by $P(Z \le z_a) = a$ for $Z \sim N(0,1)$.
We accept the null hypothesis when $|\widehat{\gamma}| \leq z_{1 - \widetilde{\alpha}/2}$; otherwise we reject the null hypothesis. 
In other words, if
$|\widehat{\gamma}| \leq z_{1 - \widetilde{\alpha}/2}$ we choose model ${\cal M}_1$; otherwise we choose model ${\cal M}_2$.

The least squares estimators of $\theta$ under the models ${\cal M}_2$ and ${\cal M}_1$ 
are 
$\widehat{\theta}$ and 
$\widehat{\theta}  - \rho \, \sigma \, v_{\theta}^{1/2} \, \widehat{\gamma}$, respectively.
Therefore the post-model-selection 
estimator of $\theta$ is
\begin{equation}
\label{mu_hat}
\widehat{\theta}_{\textsc{\tiny PMS}} =
\begin{cases}
\widehat{\theta} - \rho \, \sigma \, v_\theta^{1/2} \, \widehat{\gamma}
&\text{if}  \ \ |\widehat{\gamma}| \leq z_{1 - \widetilde{\alpha}/2}  \\
\widehat{\theta} &\text{otherwise.}
\end{cases}
\end{equation}
For $\rho \ne 0$, we note that $\widehat{\theta}_{\textsc{\tiny PMS}}$
is (in the terminology of Efron, 2014) a ``jumpy'' estimate: as $|\widehat{\gamma}|$
increases through the value $z_{1 - \widetilde{\alpha}/2}$, $\widehat{\theta}_{\textsc{\tiny PMS}}$ will change discontinuously.
Henceforth, we suppose that the known quantities $\rho$ and $v_{\theta}$
and the size $\widetilde{\alpha}$  are given.

\bigskip

\noindent {\bf 3. Computationally convenient exact formulae for the ideal bootstrap smoothed 
estimate, standard deviation and delta-method approximation to the standard  deviation}

\medskip

Efron (2014) describes the ideal bootstrap smoothed estimate $\widetilde{\theta}$ of $\theta$ by considering a limit as the number
of boostrap resamples $B \rightarrow \infty$. Because we are dealing with a parametric bootstrap, we are able to express
the ideal bootstrap smoothed estimate as follows. Let $E_{\bbeta}(\widehat{\theta}_{\textsc{\tiny PMS}})$ denote the expected value of $\widehat{\theta}_{\textsc{\tiny PMS}}$,
for true parameter value $\bbeta$.
The ideal bootstrap smoothed estimate
$\widetilde{\theta}$ is obtained by first evaluating $E_{\bbeta}(\widehat{\theta}_{\textsc{\tiny PMS}})$ and then replacing
$\bbeta$ by $\widehat{\bbeta}$.

The following theorem, proved in the appendix, provides a
computationally convenient exact formula for
$E_{\bbeta}(\widehat{\theta}_{\textsc{\tiny PMS}})$. Let $\Phi$ and $\phi$ denote the $N(0,1)$ cumulative distribution function and probability density function, respectively.
\begin{theorem}
Let $k(\gamma) = \phi(d + \gamma) - \phi(d - \gamma) + \gamma \big [\Phi(d - \gamma) - \Phi(-d-\gamma) \big]$. Then
$E_{\bbeta}(\widehat{\theta}_{\textsc{\tiny PMS}}) 
= \theta - \rho \, \sigma \, v_\theta^{1/2} \, k(\gamma)$.
Note that $k(0)=0$ and $k(\gamma)$ is an odd function of $\gamma$ that takes positive values for all $\gamma > 0$ and approaches 0 as $\gamma \rightarrow \infty$.
\end{theorem}
\noindent It follows from this theorem that the ideal bootstrap smoothed estimator $\widetilde{\theta}$ satisfies 
\begin{equation}
\label{FormulaThetaTilde}
\widetilde{\theta} = \widehat{\theta} - \rho \, \sigma \, v_\theta^{1/2} \, k(\widehat{\gamma}).
\end{equation}

\bigskip

The following theorem, proved in the appendix, provides a computationally convenient exact formula for the standard deviation of
$\widetilde{\theta}$. 
We denote this standard deviation by ${\tt sd}(\gamma)$.
\begin{theorem}
The standard deviation of $\widetilde{\theta}$ is a function of $\gamma$, which we denote by ${\tt sd}(\gamma)$, is
$\sigma \, v_{\theta}^{1/2} \, r(\gamma; \rho)$, where
\begin{equation*}
r(\gamma; \rho) =
\left( 1 - 2 \rho^2 \int_{-\infty}^{\infty} k(z) \, (z - \gamma) \, \phi(z - \gamma) \, dz
+ \rho^2 \int_{-\infty}^{\infty}\big(k(z) - m_k(\gamma))^2 \, \phi(z-\gamma) \, dz \right)^{1/2},
\end{equation*}
for
\begin{equation*}
m_k(\gamma) = \int_{-\infty}^{\infty} k(z) \, \phi(z - \gamma) \, dz.
\end{equation*}
\end{theorem}
%

The following theorem, proved in the appendix, provides a
computationally convenient exact
formula for the delta-method approximation to the standard deviation of the ideal bootstrap smoothed estimator $\widetilde{\theta}$.
\begin{theorem}
Let $q(\gamma) = \Phi(d-\gamma) - \Phi(-d-\gamma) - d \, \big[ \phi(d+\gamma) + \, \phi(d-\gamma) \big]$.
Note that $q(\gamma)$ is an even function of $\gamma$.
The delta-method approximation to the standard deviation of $\widetilde{\theta}$ is a function of $\gamma$,
which we denote by ${\tt sd}_{\rm delta}(\gamma)$, and is 
$\sigma \, v_{\theta}^{1/2} \, r_{\rm delta}(\gamma; \rho)$, where
\begin{equation*}
r_{\rm delta}(\gamma; \rho)
= \big( 1 - 2 \rho^2 q(\gamma) + \rho^2 q^2(\gamma) \big)^{1/2}.
\end{equation*}
\end{theorem}
%

\medskip

We consider the following confidence intervals for $\theta$
centered on the bootstrap smoothed esimator $\widetilde{\theta}$, with nominal 
coverage $1 - \alpha$:
\begin{align*}
J 
&= \left[ \widetilde{\theta} - z_{1 - \alpha/2} \, {\tt sd}(\widehat{\gamma}), 
\, \widetilde{\theta} + z_{1 - \alpha/2} \, {\tt sd}(\widehat{\gamma}) \right]
\qquad \ \ \ \ \ \ \ \ \ \big({\tt sd\;interval}\big)
\\
J_{\rm delta} 
&= \left[ \widetilde{\theta} - z_{1 - \alpha/2} \, {\tt sd}_{\rm delta}(\widehat{\gamma}), 
\, \widetilde{\theta} + z_{1 - \alpha/2} \, {\tt sd}_{\rm delta}(\widehat{\gamma}) \right] \qquad \big({\tt sd}_{\rm delta}\,{\tt interval} \big),
\end{align*}
%

\bigskip

\noindent {\bf 4. Coverage probability of the confidence interval centered on the bootstrap smoothed estimator}

\medskip

Let  $CP(\gamma, \rho)$ and $CP_{\, \rm{delta}}(\gamma, \rho)$
denote the coverage probabilities $P(\theta \in J)$ and $P(\theta \in J_{\, \rm{delta}})$, respectively.
Also let $\Phi(\ell, u ; \mu, v) = P(\ell \leq Z \leq u)$ for $Z \sim N(\mu, v)$. The following theorem is proved in the appendix.

\begin{theorem}
	\label{CPofJ}
	Let
$\ell(\gamma, \rho)	
= - z_{1-\alpha/2} \, r(\gamma; \rho)
	+ \rho \, k(\gamma)$
and	$u(\gamma, \rho)
= z_{1-\alpha/2} \, r(\gamma; \rho)
	+ \rho \, k(\gamma)$.
Then \newline
	(a)
	\begin{equation}
	\label{FormulaForCPofJ}
	CP(\gamma, \rho)
	= \int_{-\infty}^{\infty} \Phi \big( \ell(h, \rho), u(h, \rho); \rho(h-\gamma), 1-\rho^2 \big) \, \phi(h-\gamma) \, dh,
	\end{equation}
	(b)
 	For every given $\rho$, $CP(\gamma, \rho)$ is an even function of $\gamma$
 	and, for every given $\gamma$, $CP(\gamma, \rho)$ is an even function of $\rho$.
 \end{theorem}

The proof of the following theorem is the same as the proof of Theorem
\ref{CPofJ}, but with 
$r(\gamma; \rho)$ replaced by
$r_{\, \rm delta}(\gamma; \rho)$.

\begin{theorem}
	\label{CPofJ_delta}
	Let $\ell_{\, \rm delta}(\gamma, \rho)
	=  - z_{1-\alpha/2} \, r_{\, \rm delta}(\gamma; \rho) + \rho \, k(\gamma)$ and
	\newline
$u_{\, \rm delta}(\gamma, \rho)
=  z_{1-\alpha/2}\, r_{\, \rm delta}(\gamma; \rho) + \rho \, k(\gamma)$.
 Then \newline
	(a)
	\begin{equation*}
	CP_{\, \rm{delta}}(\gamma, \rho)
	= \int_{-\infty}^{\infty} \Phi \big( \ell_{\, \rm delta}(h, \rho),
	u_{\, \rm delta}(h, \rho);
	\rho(h-\gamma), 1-\rho^2 \big) \, \phi(h-\gamma) \, dh.
	\end{equation*}
(b)
	For every given $\rho$, $CP_{\, \rm delta}(\gamma, \rho)$ is an even function of $\gamma$
	and, for every given $\gamma$, $CP_{\, \rm delta}(\gamma, \rho)$ is an even function of $\rho$.
\end{theorem}

\medskip

\noindent {\bf 5. Scaled expected length of the confidence interval centered on the bootstrap smoothed estimator}

\medskip

The scaled expected length of the confidence interval $J$, with nominal coverage $1 - \alpha$, is defined as follows. Let $c_{\rm min}$ denote the minimum coverage probability of this confidence interval. Now let $I(c)$ denote the usual confidence interval for $\theta$, with coverage $c$, 
based on the full model. In other words, let
$I(c) = \big[ \widehat{\theta} - z_{(1+c)/2} \, \sigma \, v_{\theta}^{1/2}, 
\widehat{\theta} + z_{(1+c)/2} \, \sigma \,  v_{\theta}^{1/2}  \big]$.
The scaled expected length of $J$, denoted $SEL(\gamma, \rho)$, is defined to be the ratio 
$E(\text{length of } J) / E(\text{length of } I(c_{\rm min}))$.
The following theorem is proved in the appendix.

\begin{theorem}
	\label{SELofJ}
	Let $c_{\rm min}$ denote the minimum coverage probability of
	the confidence interval $J$, with nominal coverage $1 - \alpha$.
	Then
	
\noindent 	(a)
	\begin{equation*}
	SEL(\gamma, \rho)
	= \frac{z_{1-\alpha/2}}{z_{(1+c_{\rm min})/2}} \int_{-\infty}^{\infty} r(h;\rho) \, \phi(h-\gamma) \, dh.
	\end{equation*}
	(b)
	For every given $\rho$, $SEL(\gamma, \rho)$ is an even function of $\gamma$
	and, for every given $\gamma$, $SEL(\gamma, \rho)$ is an even function of $\rho$.
\end{theorem}

The scaled expected length of the confidence interval $J_{\rm delta}$,
denoted by $SEL_{\rm delta}(\gamma, \rho)$,  is defined in a similar way to the  
scaled expected length of $J$.
The proof of the following theorem is the same as the proof of Theorem
\ref{SELofJ}, but with 
$r(\gamma; \rho)$ replaced by
$r_{\, \rm delta}(\gamma; \rho)$.

\begin{theorem}
	\label{SELofJdelta}
	Let $c_{\rm min}$ denote the minimum coverage probability of
	the confidence interval $J_{\rm delta}$, with nominal coverage $1 - \alpha$.
	Then
	
	\noindent (a)
	\begin{equation*}
	SEL_{\rm delta}(\gamma, \rho)
	= \frac{z_{1-\alpha/2}}{z_{(1+c_{\rm min})/2}} \int_{-\infty}^{\infty} r_{\rm delta}(h;\rho) \, \phi(h-\gamma) \, dh.
	\end{equation*}
	(b)
	For every given $\rho$, $SEL_{\rm delta}(\gamma, \rho)$ is an even function of $\gamma$
	and, for every given $\gamma$, $SEL_{\rm delta}(\gamma, \rho)$ is an even function of $\rho$.
\end{theorem}

\medskip

It follows from Theorems 6 and 7 that we are able to encapsulate the scaled expected length of 
both the ${\tt sd\;interval}$ and the 
${\tt sd}_{\rm delta}\,{\tt interval}$, for all possible choices of design matrix, parameter of interest $\theta$ and parameter $\tau$ that specifies the simpler model, using only the two 
parameters $|\rho|$
and $|\gamma|$.

The bootstrap smoothed estimator is obtained by smoothing the post-model-selection estimator that results from a preliminary test of the null hypothesis that the simpler model is correct i.e. that $\gamma = 0$.
This post-model-selection estimator is usually motivated by a desire for 
good performance when the simpler model is correct. Therefore, ideally, 
both the ${\tt sd\;interval}$ and the 
${\tt sd}_{\rm delta}\,{\tt interval}$ should have a scaled expected length that is substantially less than 1 when $\gamma = 0$. In addition, ideally, these confidence intervals should have scaled expected length that (a) has maximum value that is not too much larger than 1 and (b) approaches 1 as $|\gamma|$ approaches infinity.

Figure 2 is the graph of the scaled expected length of the
CI centred on the bootstrap smoothed estimator, which is based on the post-model-selection estimator obtained after a preliminary hypothesis test,  with size 0.1, of the null hypothesis that the simpler model is correct.  This CI has
nominal coverage 0.95 and width proportional to the estimate of  
${\tt sd}_{\rm delta}$ \big(obtained by replacing $\gamma$ by $\widehat{\gamma}$
in the expression for ${\tt sd}_{\rm delta}$ \big).
We consider $|\rho| =  0.2, 0.5, 0.7$ and 0.9.
This figure provides an illustration of the following two properties of CI's 
centred on the bootstrap smoothed estimator, with width proportional to the estimate of either ${\tt sd}$ or ${\tt sd}_{\rm delta}$
\big(obtained by replacing $\gamma$ by $\widehat{\gamma}$ \big). The 
scaled expected lengths of these CI's (a) are 
either greater than 1 or only
slightly less than 1 at $\gamma = 0$ and (b) have maximum values that are increasing functions of $|\rho|$ that can be much larger than 1 for $|\rho|$ large.
These properties are established, through extensive numerical evaluation, in the Supplementary material. 
Our overall interpretation of these two properties is that the CI 
centred on the bootstrap
smoothed estimator, with width proportional to the estimated standard deviation, 
does not perform substantially better than the 
the usual 
confidence interval, with the same minimum coverage
probability, based on the full model.

\begin{figure}[h]
	\centering
	\includegraphics[width=1\textwidth]{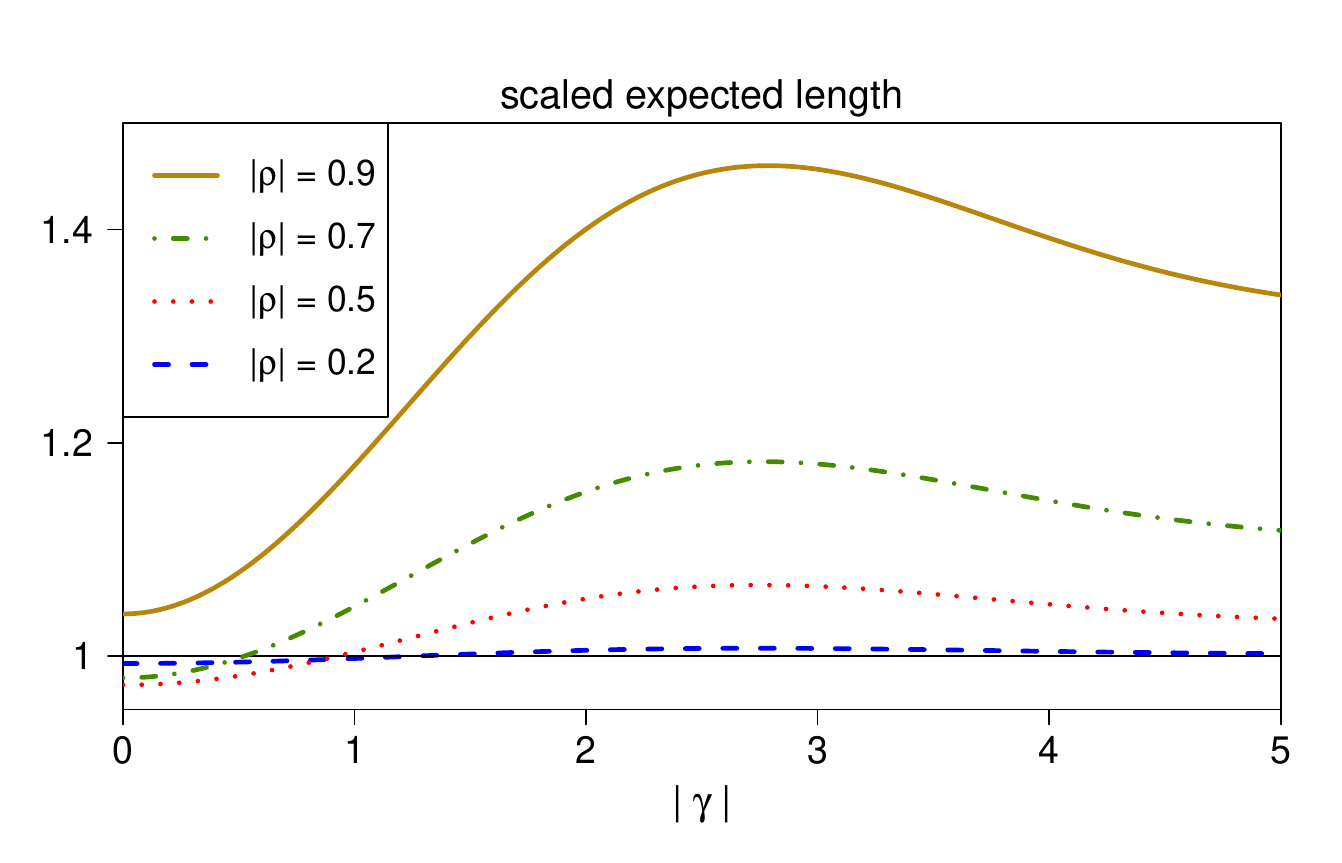}
	\caption{Graphs, for $|\rho| =  0.2, 0.5, 0.7$ and 0.9, of the scaled expected length of the
		${\tt sd}_{\rm delta}$ interval, which is based on the post-model-selection estimator obtained after a preliminary hypothesis test,  with size 0.1, of the null hypothesis that the simpler model is correct.  This CI has nominal coverage 0.95.}
\end{figure}

\bigskip

\noindent { \bf Discussion}

We have considered the scenario of two nested linear regression models, with the model chosen using a preliminary test. We have considered the case of known error variance, which is an approximation to the case that the error variance is unknown and the usual estimator of this variance is reasonably accurate. 
Also, under the appropriate large sample conditions, a logistic regression can be transformed, to a good approximation, to a linear regression model with normal errors having known error variance (see e.g. Cox, 1970, Chapter 3).

The advantage of the scenario that we consider is that we have derived computationally convenient exact expressions for all of the quantities of interest. This, in turn, has allowed us to make findings in this scenario that are valid for all design matrices, all parameters of interest that are linear combinations of the regression parameter vector, all possible preliminary tests with all possible test levels.

Usually, in practice, the bootstrap smoothed estimator is found by 
using a finite number $B$ of bootstrap resamples. This estimator is a ``noisy''
version of the ideal bootstrap smoothed estimator, which is found in the limit
as $B \rightarrow \infty$. We consider the ideal bootstrap smoothed estimator
and so we have placed the bootstrap smoothed estimator in the best possible light.

 We have considered a confidence interval, with 
nominal coverage $1 - \alpha$ and with half-width equal to the $1 - \alpha/2$ quantile of the standard normal distribution multiplied by the estimate of the 
standard deviation of this estimator. We call this the
${\tt sd}\,{\tt interval}$. 
We have also considered the same confidence
interval, but with this standard deviation
replaced by the delta method approximation to it.
We call this interval the ${\tt sd}_{\rm delta}\,{\tt interval}$. 
We have shown that both of these confidence intervals outperform the 
post-model-selection confidence interval, 
with the same nominal coverage 
and based on the same preliminary test, in terms of minimum coverage probability.

We have found, however, that the ${\tt sd}\,{\tt interval}$ and 
${\tt sd}_{\rm delta}\,{\tt interval}$ do not perform any better in terms of expected length than the usual confidence interval, with the same minimum coverage probability and based on the full model.
This is consistent with the observation by Hjort (2014) that one would expect to be able to improve on the ${\tt sd}_{\rm delta}\,{\tt interval}$ because the distribution of the difference between the bootstrap smoothed estimator and the true parameter value is ``typically highly nonnormal, asymmetric etc.''
The choice of data-based width of a confidence interval  
centred on the bootstrap smoothed estimator has a crucial role in determining the performance, in terms of the coverage and expected length, of this interval. Our conclusion is that finding a good recipe for this data-based width is still an open problem.

\bigskip



%

\noindent { \bf Appendix:  Proofs of Theorems 1, 2, 3, 4 and 6}

\medskip


In this appendix we prove Theorems 1, 2, 3, 4 and 6.
These proofs use the following lemma.
\begin{lemma}
\begin{align*}
\left[ {\begin{array}{c}
\widehat{\theta} \\
\widehat\gamma
\end{array} } \right] \sim N \left(
\left[ {\begin{array}{c}
\theta \\
\gamma
\end{array} } \right],
\left[ {\begin{array}{cc}
\sigma^2 \, v_\theta & \rho \, \sigma \, {v_\theta}^{1/2} \\
\rho \, \sigma \, {v_\theta}^{1/2} & 1
\end{array} } \right]
\right).
\end{align*}
\end{lemma}

\medskip

\noindent \textbf{Proof of Theorem 1}

\noindent To make the dependence of $\widehat{\theta}_{\textsc{\tiny PMS}}$ on $\widehat{\theta}$ and $\widehat{\gamma}$
explicit, we write $\widehat{\theta}_{\textsc{\tiny PMS}} = h(\widehat{\theta}, \widehat{\gamma})$.
Now
\begin{align*}
E_{\bbeta}(\widehat{\theta}_{\textsc{\tiny PMS}})
&= \int_{-\infty}^{\infty}
E_{\bbeta} \big( h(\widehat{\theta}, z) \, | \, \widehat{\gamma}=z \big) \phi(z-\gamma) \, dz
 \, , \quad \text{since } \ \widehat{\gamma} \sim N(\gamma, 1),  \\
&= \int_{-d}^{d} E_{\bbeta} \big( h(\widehat{\theta}, z) \, | \, \widehat{\gamma}=z \big) \phi(z-\gamma) \, dz
+ \int_{-\infty}^{-d} E_{\bbeta} \big( h(\widehat{\theta}, z) \, | \, \widehat{\gamma}=z \big) \phi(z-\gamma) \, dz \\
&\ \ \ \ \ \ \ + \int_{d}^{\infty} E_{\bbeta} \big( h(\widehat{\theta}, z) \, | \, \widehat{\gamma} = z \big) \phi(z-\gamma) \, dz \\
&= \int_{-d}^{d} E_{\bbeta} \big( \widehat\theta - \rho \, \sigma \, v_\theta^{1/2} z \, | \, \widehat{\gamma}=z \big) \phi(z-\gamma) \, dz \\
&\ \ \ \ \ \ \ + \int_{-\infty}^{-d} E_{\bbeta} \big( \widehat\theta \, | \, \widehat{\gamma}=z \big) \phi(z-\gamma) dz
+ \int_{d}^{\infty} E_{\bbeta} \big( \widehat{\theta} \, | \, \widehat{\gamma}=z \big) \phi(z-\gamma) \, dz \, , \ \text{by } \eqref{mu_hat} \text{,}
	\\
&= \int_{-\infty}^{\infty} E_{\bbeta} \big( \widehat\theta \, | \, \widehat{\gamma}=z \big) \phi(z-\gamma) \, dz
- \rho \, \sigma \, v_\theta^{1/2} \int_{-d}^{d} z\ \phi(z-\gamma) \, dz \\
&= E_{\bbeta}(\widehat{\theta}) - \rho \, \sigma \, v_\theta^{1/2} \int_{-d}^{d} z\ \phi(z-\gamma) \, dz \\
&= \theta - \rho \, \sigma \, v_\theta^{1/2} k(\gamma),
\end{align*}
where
\begin{equation*}
k(\gamma) = \int_{-d}^{d} z\ \phi(z-\gamma) \, dz.
\end{equation*}
The formula for and properties of $k(\gamma)$ stated in the theorem are proved in the Supplementary material.


\medskip

\noindent \textbf{Proof of Theorem 2}

\noindent It follows from \eqref{FormulaThetaTilde} that
\begin{align*}
\var(\widetilde{\theta})
&= \var(\widehat{\theta}) + \rho^2 \, \sigma^2 \, v_{\theta} \, \var(k(\widehat{\gamma}))
 - 2 \, \rho \, \sigma \, v_{\theta}^{1/2} \, E \big((\widehat{\theta} - \theta)(k(\widehat{\gamma}) - E(k(\widehat{\gamma}))) \big) \\
&= \sigma^2 \, v_{\theta} + \rho^2 \, \sigma^2 \, v_{\theta} \, \int_{-\infty}^{\infty} \big(k(z) - m_k(\gamma))^2 \, \phi(z-\gamma) \, dz
 - 2 \, \rho \, \sigma^2 \, v_{\theta} E \big( G \, k(\widehat{\gamma}) \big),
\end{align*}
where $G = (\widehat{\theta} - \theta) / \big( \sigma v_\theta^{1/2} \big)$.
Now
$E\big( G \, k(\widehat{\gamma}) \, | \, \widehat{\gamma}=z \big)
= k(z) \, E\left(G \, | \, \widehat{\gamma}=z \right)
= \rho \, k(z) \, (z-\gamma)$, since 
\begin{align}
\label{JointPdf_G_gammahat}
\left[ {\begin{array}{c}
	G \\
	\widehat{\gamma}
	\end{array} } \right] \sim N \left(
\left[ {\begin{array}{c}
	0 \\
	\gamma
	\end{array} } \right],
\left[ {\begin{array}{cc}
	1 & \rho  \\
	\rho  & 1
	\end{array} } \right]
\right),
\end{align}
by Lemma 1.
Thus
\begin{align*}
E \left( G \, k(\widehat\gamma)\right)
&= \rho \int_{-\infty}^{\infty} k(z) (z-\gamma) \phi(z-\gamma) dz.
\end{align*}
%


\noindent \textbf{Proof of Theorem 3}

\noindent To prove that $q$ is an even function, we need to prove that
\begin{equation*}
q(-\gamma) = \Phi(d + \gamma) - \Phi(-d + \gamma) - d \, \big [\phi(-d + \gamma) + \phi(d + \gamma) \big]
\end{equation*}
is equal to
\begin{equation*}
q(\gamma) = \Phi(d-\gamma) - \Phi(-d-\gamma) - d \, \big [\phi(-d-\gamma) + \phi(d-\gamma) \big].
\end{equation*}
Since $\Phi(z) = 1 - \Phi(-z)$, $\Phi(d-\gamma) - \Phi(-d-\gamma) = \Phi(d + \gamma) - \Phi(-d + \gamma)$.
The result follows from $\phi(-d + \gamma) + \phi(d + \gamma) = \phi(-d-\gamma) + \phi(d-\gamma)$, since
$\phi$ is an even function.

The formula for ${\tt sd}_{\text{delta}}(\gamma) $ can be derived using Theorem 2 of Efron (2014). However, in the present scenario, the same formula results from the 
application of the delta-method approximation that uses the first order Taylor expansion,
$k(\widehat{\gamma}) \approx k(\gamma) + k^{\prime}(\gamma) \big(\widehat{\gamma} - \gamma \big)$.
It follows from \eqref{FormulaThetaTilde} that
\begin{equation*}
\widetilde{\theta} \approx \widehat{\theta} - \rho \, \sigma \, v_\theta^{1/2} \,
\big(k(\gamma) + k^{\prime}(\gamma) \big(\widehat{\gamma} - \gamma \big) \big).
\end{equation*}
The variance of the right-hand side is $\sigma^2  v_{\theta} \big(1 - 2 \rho^2 k^{\prime}(\gamma) + \rho^2 (k^{\prime}(\gamma))^2 \big)$.
Using the definition of Hermite polynomials, it may be shown that $k^{\prime}(\gamma) = q(\gamma)$.

\bigskip

\noindent \textbf{Proof of Theorem 4}
\newline
\noindent \textbf{Part (a)}
\begin{align*}
P\left( \theta \in J \right)
&= P\left( - z_{1-\alpha/2} \, \sd(\widehat\gamma) \leq \widetilde\theta - \theta \leq  z_{1-\alpha/2} \, \sd(\widehat\gamma) \right)
\\
&= P\left( - z_{1-\alpha/2} \, \sd(\widehat\gamma) \leq \widehat{\theta} - \theta -\rho \, \sigma \, {v_\theta}^{1/2} k(\widehat\gamma) \leq  z_{1-\alpha/2} \, \sd(\widehat\gamma) \right) \, ,
\ \text{by } \eqref{FormulaThetaTilde},
\\
&= P\left( - z_{1-\alpha/2} \frac{\sd(\widehat\gamma)}{\sigma \,{v_\theta}^{1/2}} 
\leq G -\rho  k(\widehat\gamma) 
\leq  z_{1-\alpha/2} \frac{\sd(\widehat\gamma)}{\sigma \,{v_\theta}^{1/2}} \right) 
\, , \ \text{where}\ G = (\widehat{\theta} -\theta)/\big( \sigma \, {v_\theta}^{1/2} \big),
\\
&= P\left( \ell(\widehat{\gamma}, \rho) \leq G  \leq  u(\widehat{\gamma}, \rho) \right)
\\
&= \int_{-\infty}^{\infty} P\left( \ell(h, \rho) \leq G \leq u(h, \rho)\, | \, \widehat\gamma=h \right) \phi(h-\gamma) \, dh.
\end{align*}
It follows from \eqref{JointPdf_G_gammahat} that
the distribution of $G$ conditional on $\widehat\gamma=h$ is $N\left( \rho(h-\gamma), 1-\rho^2\right)$.
Hence
\begin{equation*}
\int_{-\infty}^{\infty} P\left( \ell(h, \rho) \leq G \leq u(h, \rho)|\widehat\gamma=h \right) \phi(h-\gamma) dh
= \int_{-\infty}^{\infty} P\left( \ell(h, \rho) \leq \widetilde{G} \leq u(h, \rho) \right) \phi(h-\gamma) dh
\end{equation*}
where $\widetilde{G} \sim N\left( \rho(h-\gamma), 1-\rho^2\right)$.
Therefore \eqref{FormulaForCPofJ} holds.

\medskip

\noindent \textbf{Part (b):} \ 
 Our proof will use the following easily-established lemmas.

\begin{lemma}
	\label{PropertyOfPhiabmv}
	$\Phi(\ell,u; \mu,v) = \Phi(-u,-\ell; -\mu,v)$.
\end{lemma}


\begin{lemma}
	\label{PropertyOfab}
	(a) $-u(-x, \rho) = \ell(x, \rho)$. \newline
	(b) $\ell(x, -\rho) = \ell(-x, \rho)$
	and $u(x, -\rho) = u(-x, \rho)$.
\end{lemma}

%
%
%
%

Firstly, we prove that, for every given $\rho$, $CP(\gamma, \rho)$ is an even function of $\gamma$.
By Lemma \ref{PropertyOfPhiabmv} and since $\phi$ is an even function,
\begin{align*}
CP(-\gamma, \rho)
&= \int_{-\infty}^{\infty}
\Phi\big( -u(h,\rho), -\ell(h, \rho); \rho(-h-\gamma), 1-\rho^2 \big) \phi(-h-\gamma) \, dh
\\
&= \int_{-\infty}^{\infty}
\Phi\big( -u(-x,\rho), -\ell(-x, \rho); \rho(x-\gamma), 1-\rho^2 \big) \phi(x-\gamma) \, dx
\\
&\qquad \qquad (\text{by changing the variable of integration to}\ x = - h)
\\
&= \int_{-\infty}^{\infty}
\Phi\big( \ell(x,\rho), u(x, \rho); \rho(x-\gamma), 1-\rho^2 \big) \phi(x-\gamma) \, dx \, ,
\ \text{by Lemma \ref{PropertyOfab} (a)},
\\
&= CP(\gamma,\rho)
\end{align*}

We now prove that, for every given $\gamma$, $CP(\gamma, \rho)$ is an even function of $\rho$.
 Now
\begin{align*}
CP(\gamma, -\rho) &= \int_{-\infty}^{\infty} \Phi\left( \ell(h, -\rho), u(h, -\rho); -\rho(h-\gamma), 1-\rho^2 \right) \, \phi(h-\gamma) \, dh \\
&= \int_{-\infty}^{\infty} \Phi\left( -u(h, -\rho),-\ell(h, -\rho) ; \rho(h-\gamma), 1-\rho^2 \right) \, \phi(h-\gamma) \, dh \, , \ \text{by Lemma \ref{PropertyOfPhiabmv},} \\
&= \int_{-\infty}^{\infty} \Phi\left( \ell(h, \rho),u(h, \rho) ; \rho(h-\gamma), 1-\rho^2 \right) \, \phi(h-\gamma) \, dh \, , \ \text{by Lemma \ref{PropertyOfab},} \\
&= CP(\gamma, \rho)
\end{align*}

\bigskip

\noindent \textbf{Proof of Theorem 6}

\noindent \textbf{Part (a):}
\
By Theorem 2, the length of the confidence interval $J$, with nominal coverage $1 - \alpha$,
is $2 z_{1 - \alpha/2} \, \sigma \, v_{\theta}^{1/2} r(\widehat{\gamma}; \rho)$.
Thus the expected length of this CI is 
$2 z_{1 - \alpha/2} \, \sigma \, v_{\theta}^{1/2} E(r(\widehat{\gamma}, \rho))$.
Also, the length of $I(c_{\rm min})$ is 
$2 \, z_{(1+c_{\rm min})/2} \, \sigma \, v_{\theta}^{1/2}$. Thus
\begin{equation*}
SEL(\gamma, \rho)
= \frac{z_{1-\alpha/2}}{z_{(1+c_{\rm min})/2}} E(r(\widehat{\gamma}, \rho))
= \frac{z_{1-\alpha/2}}{z_{(1+c_{\rm min})/2}} \int_{-\infty}^{\infty} r(h;\rho) \, \phi(h-\gamma) \, dh.
\end{equation*}

\noindent \textbf{Part (b):} Our proof will use the following lemma.

\begin{lemma}
For every given $\rho$, $r(\gamma; \rho)$ is an even function of $\gamma$
and, for every given $\gamma$, $r(\gamma; \rho)$ is an even function of $\rho$.
\end{lemma}
Since $\phi$ is an even function,
\begin{align*}
SEL(-\gamma, \rho)
&= \frac{z_{1-\alpha/2}}{z_{(1+c_{\rm min})/2}} \int_{-\infty}^{\infty} r(h;\rho) \, \phi(-h-\gamma) \, dh
\\
&= \frac{z_{1-\alpha/2}}{z_{(1+c_{\rm min})/2}} \int_{-\infty}^{\infty} r(-x;\rho) \, \phi(x-\gamma) \, dx
\\
&\qquad \qquad (\text{by changing the variable of integration to}\ x = - h)
\\
&= SEL(\gamma, \rho),
\end{align*}
by Lemma 4. It also follows directly from this lemma that $SEL(\gamma, \rho)$
 is an even function of $\rho$, for every given $\gamma$.

 \medskip

\noindent {\bf References}

	
\smallskip
	
\rf Breiman, L. (1996). Bagging predictors. \textsl{Machine Learning} 24, 123--140

\smallskip

\rf Buckland, S.T., Burnham, K.P. and Augustin, N.H. (1997). Model selection: an integral
part of inference. \textsl{Biometrics} \textbf{53}, 603--618.

\smallskip

\rf Cox, D.R. (1970). The Analysis of Binary Data. Methuen, London.

\smallskip

%

\rf Efron, B. (2014). Estimation and accuracy after model selection. \textsl{Journal of the American Statistical Association} \textbf{109}, 991--1022.

\smallskip

\rf Fletcher, D., Turek, D. (2011). Model-averaged profile likelihood intervals. \textsl{Journal of Agricultural, Biological and Environmental Statistics} \textbf{17}, 38--51.

\smallskip

\rf Hjort, N.L. (2014). Comment on `Estimation and accuracy after model selection'
by B. Efron. \textsl{Journal of the American Statistical Association} \textbf{109}, 1017--1020.

\smallskip

\rf Hjort, N.L. and Claeskens, G. (2003). Frequentist model average estimators. 
\textsl{Journal of the American Statistical Association} \textbf{98}, 879--899.

\smallskip

\rf Kabaila, P. (2009). The coverage properties of confidence regions after model selection. 
\textsl{International Statistical Review} \textbf{77}, 405--414.

\smallskip

\rf Kabaila, P. (2016). The finite sample performance of the two-stage analysis
of a two-period crossover trial. \textsl{Statistics and Probability Letters}, 
117, 118--127.

\smallskip

\rf Kabaila, P., Welsh, A.H. and Abeysekera, W. (2016). Model-averaged confidence intervals. 
\textsl{Scandinavian Journal of Statistics} \textbf{43}, 35--48.

\newpage

\rf Kabaila, P., Welsh, A.H. and Mainzer, R. (2016). The performance of model averaged tail area confidence intervals. 
\textsl{Communications in Statistics - Theory and Methods}, \textbf{46}, 10718--10732.
\smallskip

\rf Leeb, H. and P\"otscher, B.M. (2005). Model selection and inference:
facts and fiction. \textsl{Econometric Theory}  21, 21--59.

\smallskip

\rf Turek, D. and Fletcher, D. (2012). Model-averaged Wald confidence intervals. \textsl{Computational Statistics and Data Analysis} \textbf{56},
2809--2815.

\smallskip

\rf Wang, L., Sherwood, B. and Li, R. (2014). Comment on `Estimation and accuracy after model selection'
by B. Efron. \textsl{Journal of the American Statistical Association} \textbf{109}, 1007--1010.


\end{document}